\documentstyle[11pt,aaspp]{article}

\newcommand{\etal}{et al.}

\newcommand{\lya}{Ly$\alpha$\ }

\def\cf{{\it cf.\/} }
\def\be{\begin{equation}}
\def\ee{\end{equation}}
\def\bdm{\begin{displaymath}}
\def\edm{\end{displaymath}}
\def\ea{{et al.\/} }

\def\kmpersec{km s$^{-1}$}
\def\NHI{N_{\rm HI}}
\def\dla{DLA}
\def\lta{\mathrel{\rlap{\lower 3pt\hbox{$\mathchar"218$}}
    \raise 2.0pt\hbox{$\mathchar"13C$}}}

\newbox\grsign \setbox\grsign=\hbox{$>$} \newdimen\grdimen \grdimen=\ht\grsign
\newbox\simlessbox \newbox\simgreatbox
\setbox\simgreatbox=\hbox{\raise.5ex\hbox{$>$}\llap
     {\lower.5ex\hbox{$\sim$}}}\ht1=\grdimen\dp1=0pt
\setbox\simlessbox=\hbox{\raise.5ex\hbox{$<$}\llap
     {\lower.5ex\hbox{$\sim$}}}\ht2=\grdimen\dp2=0pt
\newcommand{\simgt}{\mathrel{\copy\simgreatbox}}
\newcommand{\simlt}{\mathrel{\copy\simlessbox}}

\newcommand\cdunits{{\rm cm}^{-2}}
\newcommand\vunits{{\rm km}\;{\rm s}^{-1}}

\begin{document}

\title{THE POPULATION OF DAMPED LYMAN-ALPHA AND LYMAN LIMIT SYSTEMS IN THE 
          COLD DARK MATTER MODEL}

\author{Jeffrey P. Gardner$^1$, Neal Katz$^1$, Lars Hernquist$^{2,3}$,  David H. Weinberg$^4$}
\affil{E-mail:  gardner@astro.washington.edu, nsk@astro.washington.edu,
 lars@ucolick.org, dhw@payne.mps.ohio-state.edu}
\footnotetext[1]
{University of Washington, Department of Astronomy, Seattle, WA 98195}
\footnotetext[2]
{University of California, Lick Observatory, Santa Cruz, CA 95064}
\footnotetext[3]
{Presidential Faculty Fellow}
\footnotetext[4]
{Ohio State University, Department of Astronomy, Columbus, OH 43210}

\begin{abstract}

Lyman limit and damped \lya absorption systems 
probe the distribution of collapsed, cold gas at high redshift.
Numerical simulations that incorporate gravity and gas dynamics
can predict the abundance of such absorbers in cosmological models.
We develop a semi-analytical method to correct the numerical predictions
for the contribution of unresolved low mass halos, and we apply
this method to the Katz \ea (1996) simulation of the standard
cold dark matter model ($\Omega=1$, $h=0.5$, $\Omega_b=0.05$,
$\sigma_8=0.7$).  Using this simulation and higher resolution 
simulations of individual low mass systems, we determine the
relation between a halo's circular velocity $v_c$ and its
cross section for producing Lyman limit or damped \lya absorption.
We combine this relation with the Press-Schechter formula for the
abundance of halos --- itself calibrated against the simulated halo
population --- to compute the number of absorbers per unit redshift.
The resolution correction increases the predicted abundances by
about a factor of two at $z=2$, 3, and 4, bringing the predicted
number of damped \lya absorbers into quite good agreement with 
observations.  Roughly half of these systems reside in halos with
circular velocities 
$v_c \geq 100\;\vunits$ and half in halos with 
$35\;\vunits \leq v_c \leq 100\;\vunits$.  Halos with $v_c > 150 \;\vunits$
typically harbor two or more systems capable of producing damped
absorption.  Even with the resolution correction, 
the predicted abundance of Lyman limit systems is a factor
of three below observational estimates, signifying either a failure
of the standard cold dark matter model or a failure of these
simulations to resolve most of the systems responsible for Lyman limit
absorption.  By comparing simulations with and without star formation,
we find that depletion of the gas supply by star formation
affects absorption line statistics at $z\geq 2$ only for column
densities exceeding $\NHI=10^{22}\;\cdunits$, even though half of
the cold, collapsed gas has been converted to stars by $z=2$.

\end{abstract}


\section{Introduction}

Systems producing absorption in the spectra of distant quasars offer
an excellent probe of the early Universe.  At high redshifts, they
easily outnumber other observed tracers of cosmic structure, including
both normal and active galaxies.  Mounting evidence that
the high column density absorbers are young galaxies links relatively
pristine baryonic matter to highly evolved objects at the present day.
The amount of atomic hydrogen in damped \lya (\dla) absorbers
at $z \sim 3$ is comparable to the mass in stars at the current epoch (Wolfe
1988), and two \dla\ systems are known to have radial extents $\simgt
10 h^{-1}$ kpc (Briggs \etal\ 1989; Wolfe \etal\ 1993).  Photometry of
damped absorbers supports the view that they are high-redshift
galaxies (Djorgovski \etal\ 1996; Fontana \etal\ 1996).
At somewhat lower column densities and redshifts, 
deep imaging and spectroscopy indicate that Lyman
limit systems are associated with lines of sight passing near bright
galaxies (Yanny 1990; Lanzetta \& Bowen 1990; Bergeron \& Boiss\'e
1991; Steidel, Dickinson, \& Persson 1994) or galaxy clusters
(Lanzetta, Webb, \& Barcons 1996).

The interpretation of quasar absorption systems has undergone
something of a revolution during the past two years, with the
recognition that they may be gas aggregating into nonlinear structures
in hierarchical models like those invoked to account for the observed galaxy
distribution (e.g., Cen \ea 1994; Petitjean, Mucket, \&
Kates 1995; Zhang, Anninos, \& Norman 1995; Hernquist \etal\ 
1996; Miralda-Escud\'e \etal\ 1996).  In particular, Katz \etal\ 
(1996; hereafter KWHM) used
simulations that evolve baryons and dark matter in the presence of a
background radiation field to show that high column density absorbers
arise naturally in a cold dark matter (CDM) universe from radiatively
cooled gas in galaxy-sized halos, supporting the notion that damped
\lya systems are a byproduct of galaxy formation.  Together with the
results of Hernquist \etal\ (1996) for the \lya forest, the column
density distribution predicted by KWHM matches existing data
reasonably well, but it falls below the observations by factors $\approx
2$ and $\approx 8$ for \dla\ and Lyman limit absorbers, respectively.

This discrepancy can be attributed at least
partly to resolution effects in the simulations.  Owing to
computational expense, the KWHM simulation could not resolve halos
with circular velocities below $v_c \approx 100$ \kmpersec.  However,
higher resolution simulations of localized regions by Quinn, Katz, \&
Efstathiou (1996; hereafter QKE) 
indicate that halos down to $v_c \approx 35$
\kmpersec\ can host damped absorbers, so clearly the number of high
column density systems found by KWHM is artificially depressed
by the finite resolution of their simulation.

In this paper,
we overcome this numerical limitation using a two-step correction
procedure.  First, we employ the Press \& Schechter (1974) algorithm
to correct the KWHM data by extending the halo mass function to
values lower than could be resolved by their simulation.  Then, we
account for absorption by gas in these halos from a relation between
the absorption cross section for a halo and its circular velocity.
This relation is established by fitting both the KWHM data and
high-resolution simulations that use the QKE initial conditions
and the KWHM background radiation field.
These additional simulations examine localized
regions around low mass objects with sufficient resolution to resolve
halos down to $v_c \approx 35$ \kmpersec. 
Heating by the UV background prevents the collapse and cooling of gas
in smaller halos (QKE; Thoul \& Weinberg 1996).
The high-resolution volumes are small and
were chosen in a non-random way, so they cannot be used directly to
infer the number of \dla\ and Lyman limit systems.  By convolving the
absorption area vs.\ circular velocity relation with the halo mass function
given by the Press-Schechter method, we can predict the absorption at
any mass scale, effectively extending the dynamic range of the
simulations down to the lowest mass objects that produce high
column density absorption.
We also present another calculation, similar
to that in KWHM but including star formation, to quantify the effects
of gas depletion on high column density absorption.

\section{Simulations and Methods}
\label{secSimulation}

Our primary simulation, the same as that used by KWHM, follows the
evolution of a periodic cube whose edges measure 22.22 Mpc in comoving
units.  This region was drawn randomly from a CDM universe with
$\Omega=1$, $h \equiv H_0/100$ \kmpersec\ Mpc$^{-1}=0.5$, baryon
density $\Omega_b=0.05$, and power spectrum normalization
$\sigma_8=0.7$.  A uniform background radiation field was imposed to
mimic the expected ultraviolet (UV) output of quasars, with a spectrum
of the form $J(\nu) = J_0(\nu_0/\nu) F(z)$, where $\nu_0$ is the
Lyman-limit frequency, $J_0=10^{-22}$ erg s$^{-1}$ cm$^{-2}$ sr$^{-1}$
Hz$^{-1}$, and $F(z)=0$ if $z>6$, $F(z)=4/(1+z)$ if $3 \le z \le 6$,
and $F(z)=1$ if $2<z<3$.  The simulations employ $64^3$ gas and $64^3$
dark-matter particles, with a gravitational softening length of 20
comoving kpc (13 comoving kpc equivalent Plummer softening).  The
particle mass is $1.45 \times 10^8 M_\odot$ and $2.8 \times 10^9
M_\odot$ for gas and dark matter, respectively.  Detailed descriptions
of the simulation code and the radiation physics can be found in
Hernquist \& Katz (1989) and Katz, Weinberg, \& Hernquist (1996;
hereafter KWH).  The low column density absorption in this simulation
is discussed by Hernquist \etal\ (1996), and the galaxy population is
discussed by Weinberg, Hernquist, \& Katz (1996).

We also employ two simulations that have the same initial conditions,
cosmological parameters, and numerical parameters as QKE but the UV
background spectrum given above.  These comprise smaller, 10 Mpc
periodic volumes (with $\Omega=1$, $h=0.5$, $\Omega_b=0.05$ as
before), which are evolved using a hierarchical grid of particles in
the initial conditions.  The central region forms a collapsed object
that is represented using a large number of low mass particles, while
regions further away are modeled using a small number of more massive
particles.  A simulation of the same volume as QKE would require
$256^3$ particles of each species to match the resolution of the
central region throughout; the nesting technique allows us to achieve
high-resolution locally while preserving the cosmological context of
the calculation.

QKE find that a photoionizing background suppresses the collapse
and cooling of gas in halos with circular velocities 
$v_c \lta 35$ \kmpersec.  Thoul \& Weinberg (1996) find a similar
cutoff in much higher resolution, spherically symmetric calculations.
Hence, it should be possible to estimate the amount of gas capable of
producing \dla\ and Lyman limit absorption by accounting for
halos down to this cutoff in $v_c$.
Both QKE and Thoul \& Weinberg (1996) find that photoionization
has little effect on the amount of gas that cools in halos
with $v_c \simgt 60$ \kmpersec, consistent with the results of
Navarro \& Steinmetz (1996) and Weinberg \etal\ (1996).

The current generation of hydrodynamic simulations lacks
the dynamic range necessary to represent halos over the entire range
$35 < v_c \lta 300$ \kmpersec.  To overcome this limitation, we use
the approximation developed by Press \& Schechter (1974), who give the
following analytic estimate for the number density of halos of mass
$M$ at redshift $z$:
\be
	N(M,z) dM = \sqrt{2\over \pi} {\rho_0\over M} 
	 {\delta_c\over \sigma_0} \left({\gamma R_f\over R_*}\right)^2
	 \exp{\left({-\delta_c^2\over 2\sigma_0^2}\right)} dM ,
\label{PSnumber}
\ee
where $\rho_0$ is the mean comoving density, $R_f$ is the Gaussian
filter radius corresponding to mass
$M= (2\pi)^{3/2} \rho_0 R_f^3$, and $\delta_c$ is
the critical linear density contrast that corresponds to
gravitational collapse.  The
parameters $\sigma_0$, $\gamma$ and $R_*$ are related to moments of
the power spectrum (Bardeen \ea 1986).  Equation (\ref{PSnumber}) can
be integrated from $M$ to infinity to yield the number density
of objects above a given mass.
In what follows, for comparison with our simulations, we use the CDM
transfer function given by Bardeen \ea (1986).

To determine the number of \dla\ and Lyman limit systems per
unit redshift, we first fix the parameters in the Press-Schechter
algorithm so that it reproduces the mass function of our 22.22 Mpc
simulations.  Then, we use the 22.22 Mpc and 10 Mpc 
simulations together to fit a relation between the circular velocity
of a halo and its cross section for producing \dla\ 
or Lyman limit absorption.

To identify halos in the simulations, we apply a friends-of-friends
algorithm with a linking length equal to the mean interparticle
separation on an isodensity contour of an isothermal sphere with an
overdensity $177$, $b= (177 n/3)^{-1/3}$ where $n$ is the particle number
density.
We also apply the algorithm of Stadel \ea (1996;
see also KWH and
http://www-hpcc.astro.washington.edu/tools/DENMAX)
to the cold gas particles
in the same volume to locate regions of collapsed gas capable of
producing Lyman limit and damped \lya absorption.  A region of gas is
considered a potential absorber only if it contains at least four
gas particles that are mutually bound, have a smoothed overdensity
$\rho_g/\bar\rho_g > 170$, and a temperature $T < 30,000$ K.  
All of the gas concentrations found by this method
are associated with a friends-of-friends halo,
even at $z=4$.  We match each absorber with its
parent halo and discard halos that contain no absorbers.

For each of the halos that contains a cold gas concentration, we
determine the radius of the sphere centered on the most tightly bound
particle within which the average density is equal to 177 times the
mean background density.  We characterize halo masses and circular
velocities by their values at this radius.  This method of quantifying
the properties of halos in the simulations corresponds to that
envisioned in the Press-Schechter approximation, which is based on the
spherical collapse model.  We find that the mass distribution of halos
in the simulations is best fit using the Press-Schechter form with a
Gaussian filter and $\delta_c = 1.69$.  Many workers have instead used
a top-hat filter, with $M_f=(4 \pi/3) \rho_0 R_f^3$ (\cf Ma 1996; Ma
\& Bertschinger 1994; Mo \& Miralda-Escud\'e 1994; Mo \ea 1996), or a
Gaussian filter with a modified relation between filter radius and
associated mass, $M_f=6 \pi^2 \rho_0 R_f^3$ (Lacey \& Cole 1994), with
similar values for $\delta_c$.  However, these studies used the halo
masses as returned by the friends-of-friends algorithm itself, and if
we do this we also find that top-hat or modified Gaussian filters
provide good fits to the mass function for $\delta_c \approx 1.7$.
The combination $\delta_c=1.69$, Gaussian filter, and
$M_f=(2\pi)^{3/2} \rho_0 R_f^3$ is appropriate for our definition of
halo masses within overdensity 177 spheres.  Including or excluding
the ``absorberless'' halos in our mass function does not change the
results above $v_c=100$ \kmpersec\ because all halos above this
circular velocity contain at least one absorber.

We calculate HI column densities for the halos by encompassing each
halo with a sphere which is centered on the most tightly bound
gas particle and is of a sufficient size to contain all gas particles
which may contribute to absorption within the halo.  We
project the gas
distribution within this sphere onto a uniform grid of cell size 5.43
comoving kpc, equal to the highest resolution achieved anywhere in the
22.22 Mpc simulation.  
Using the method of KWHM, we calculate an initial HI column
density for each gridpoint assuming that the gas is optically thin,
then apply a self-shielding correction to yield a true HI column
density (see KWHM for details).  For each halo we compute the
projected area over which it produces damped absorption, with $\NHI >
10^{20.3} \;\cdunits$, and Lyman limit absorption, with $\NHI >
10^{17.2}\;\cdunits$.  For simplicity, we project all halos from a
single direction, though we obtain a similar fit of absorption area to
circular velocity if we project randomly in the $x$, $y$, and $z$
directions or average the projections in $x$, $y$, and $z$.

\begin{figure}
\vglue-0.65in
\plottwo{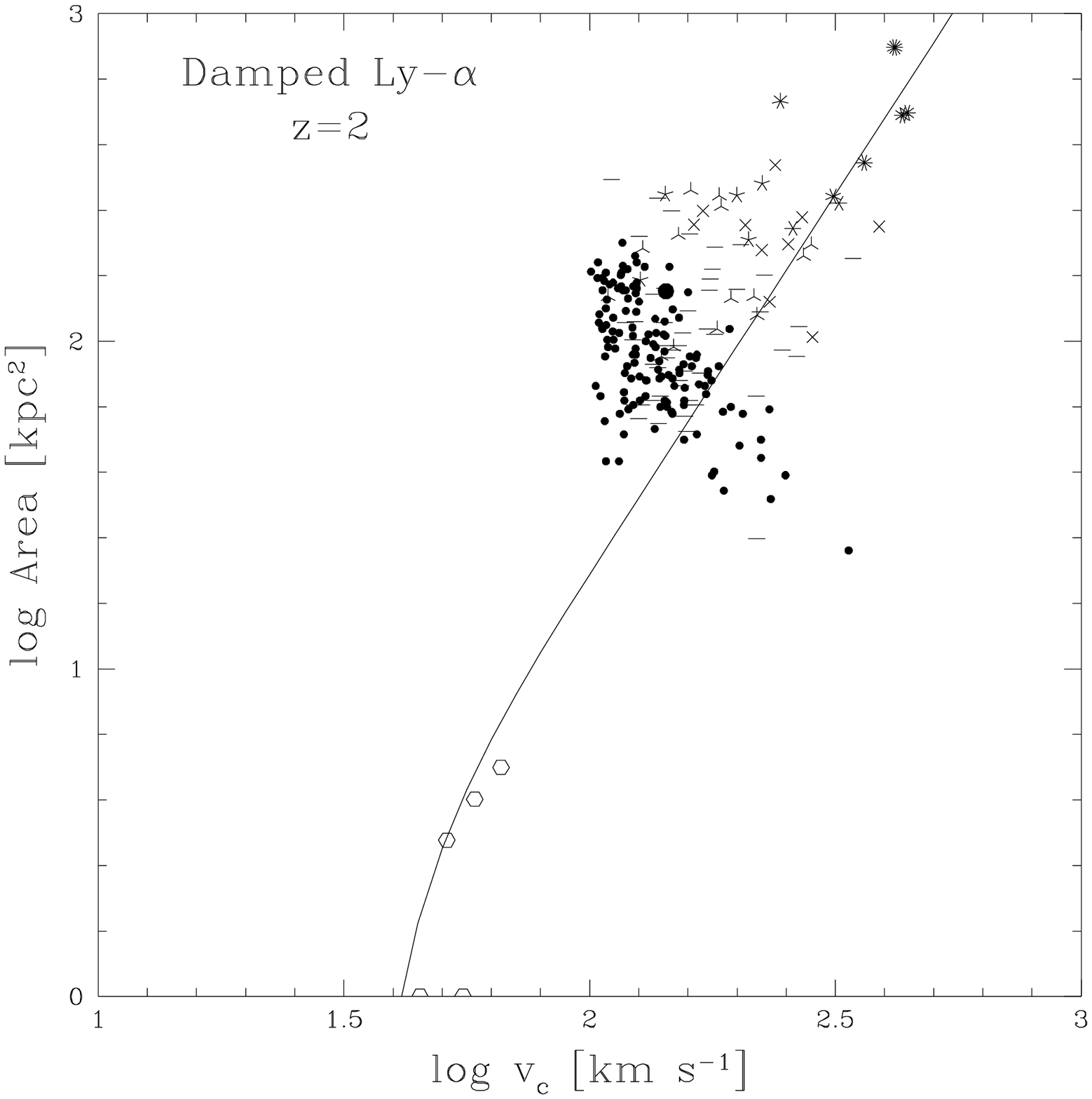}{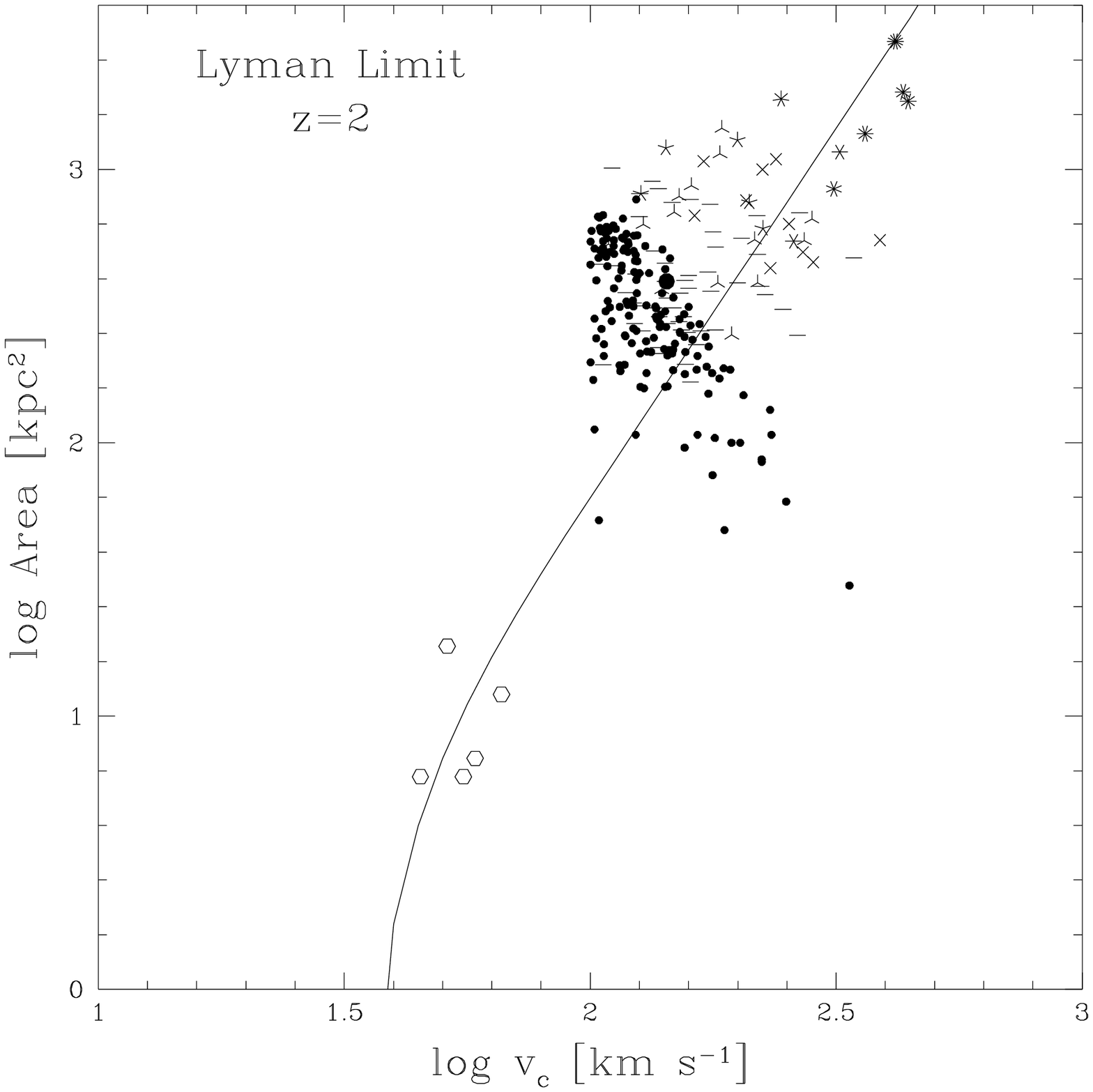} \\
\vglue-0.2in
\plottwo{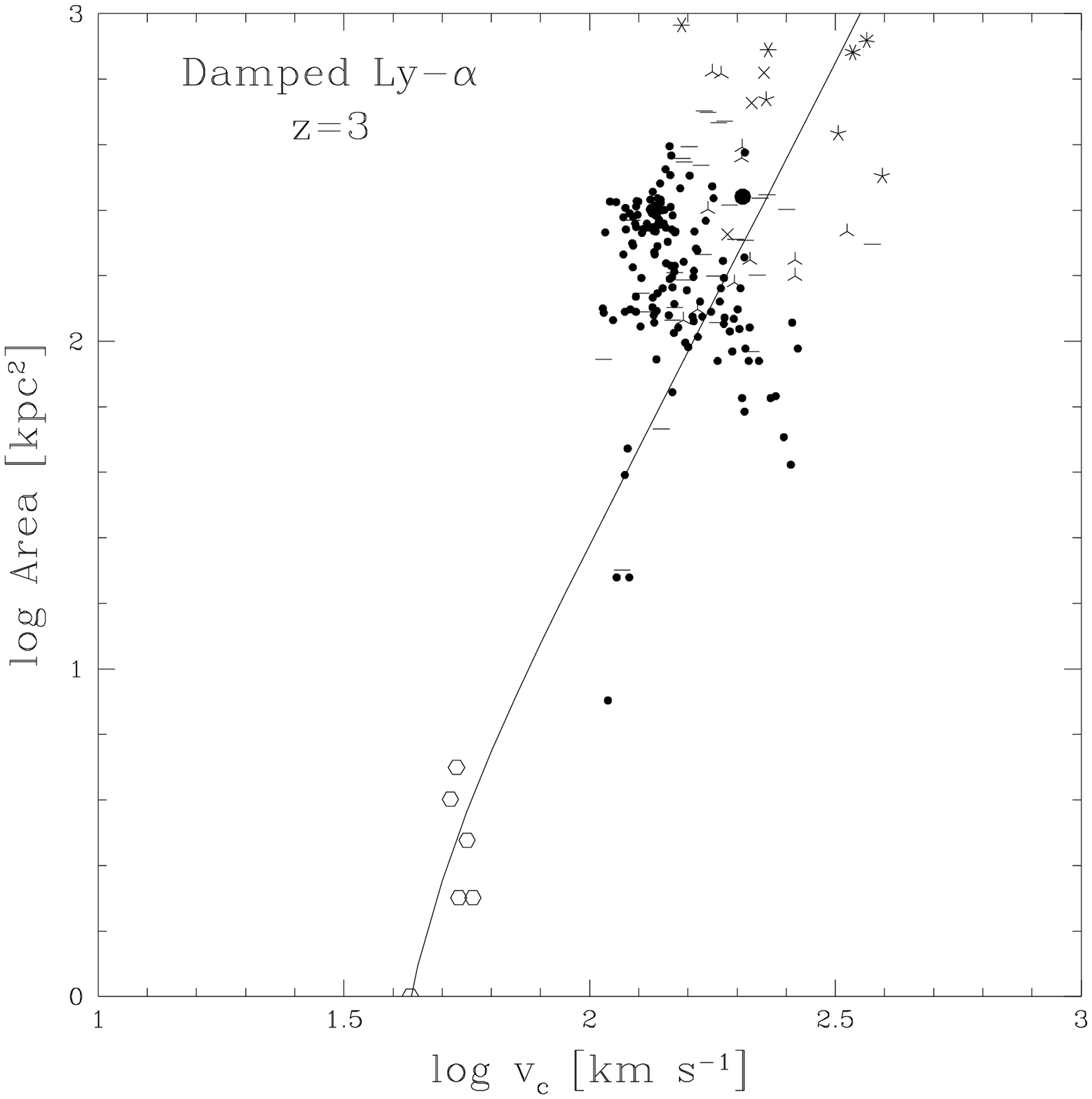}{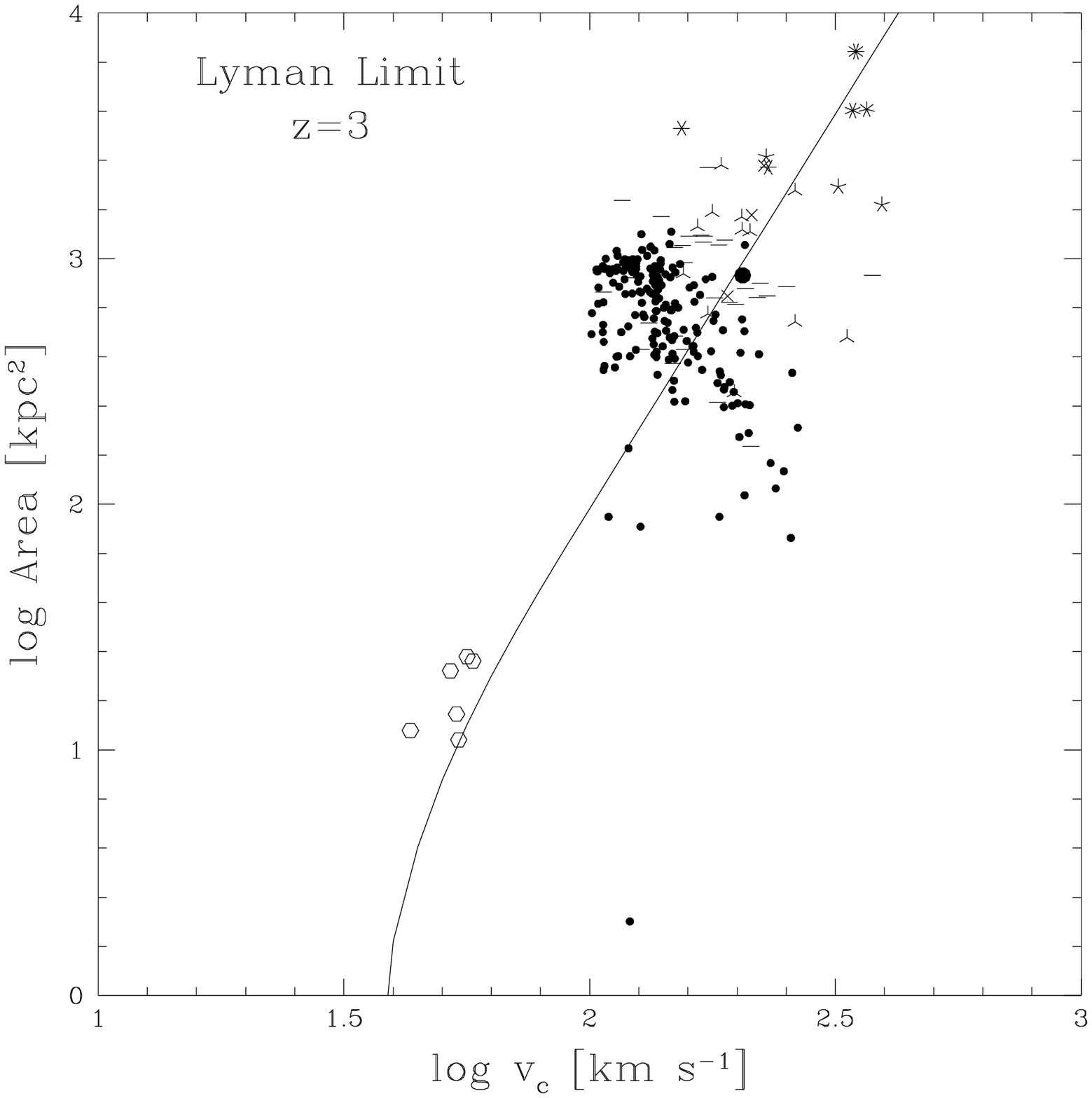} \\
\vglue-0.2in
\plottwo{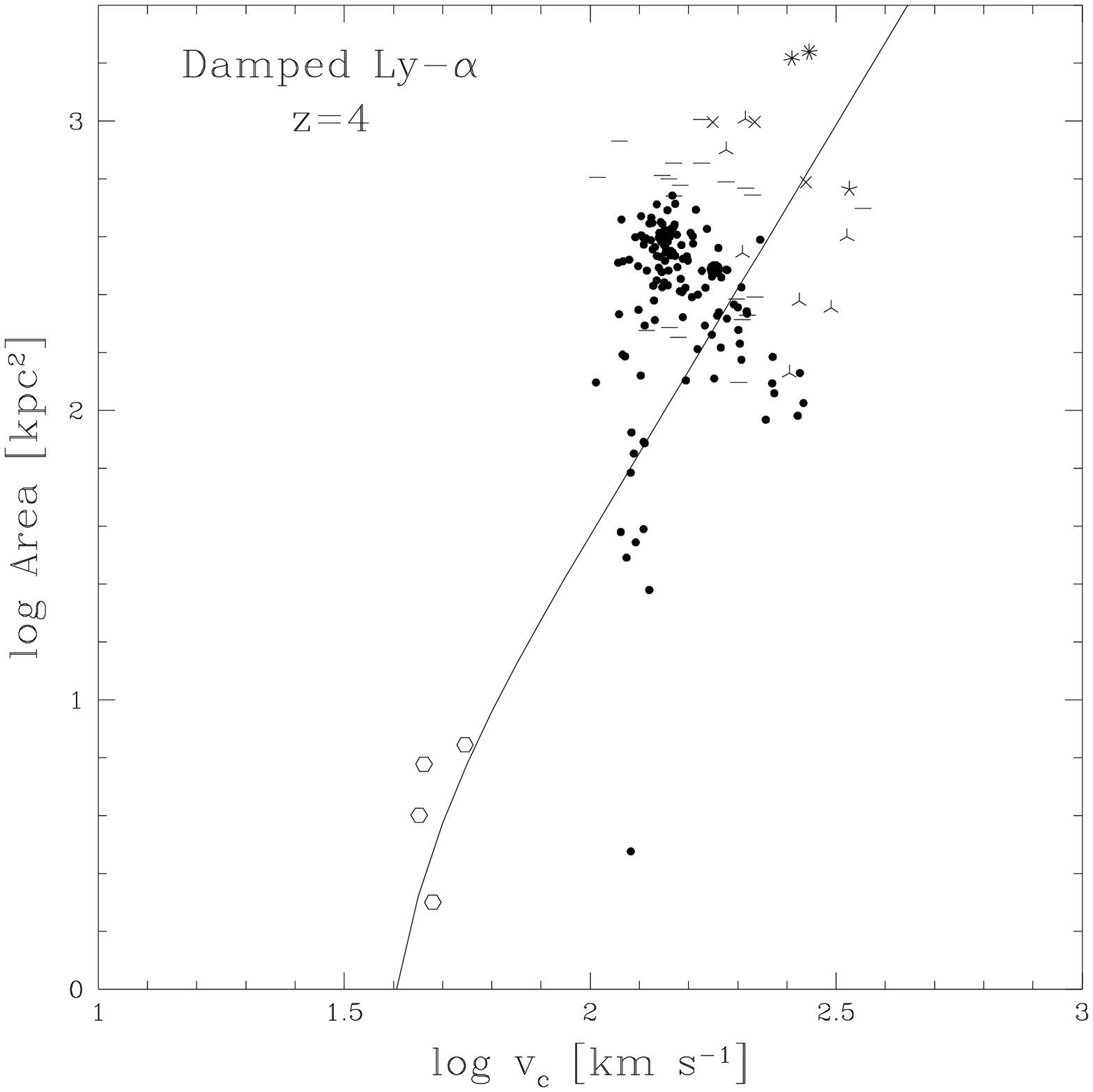}{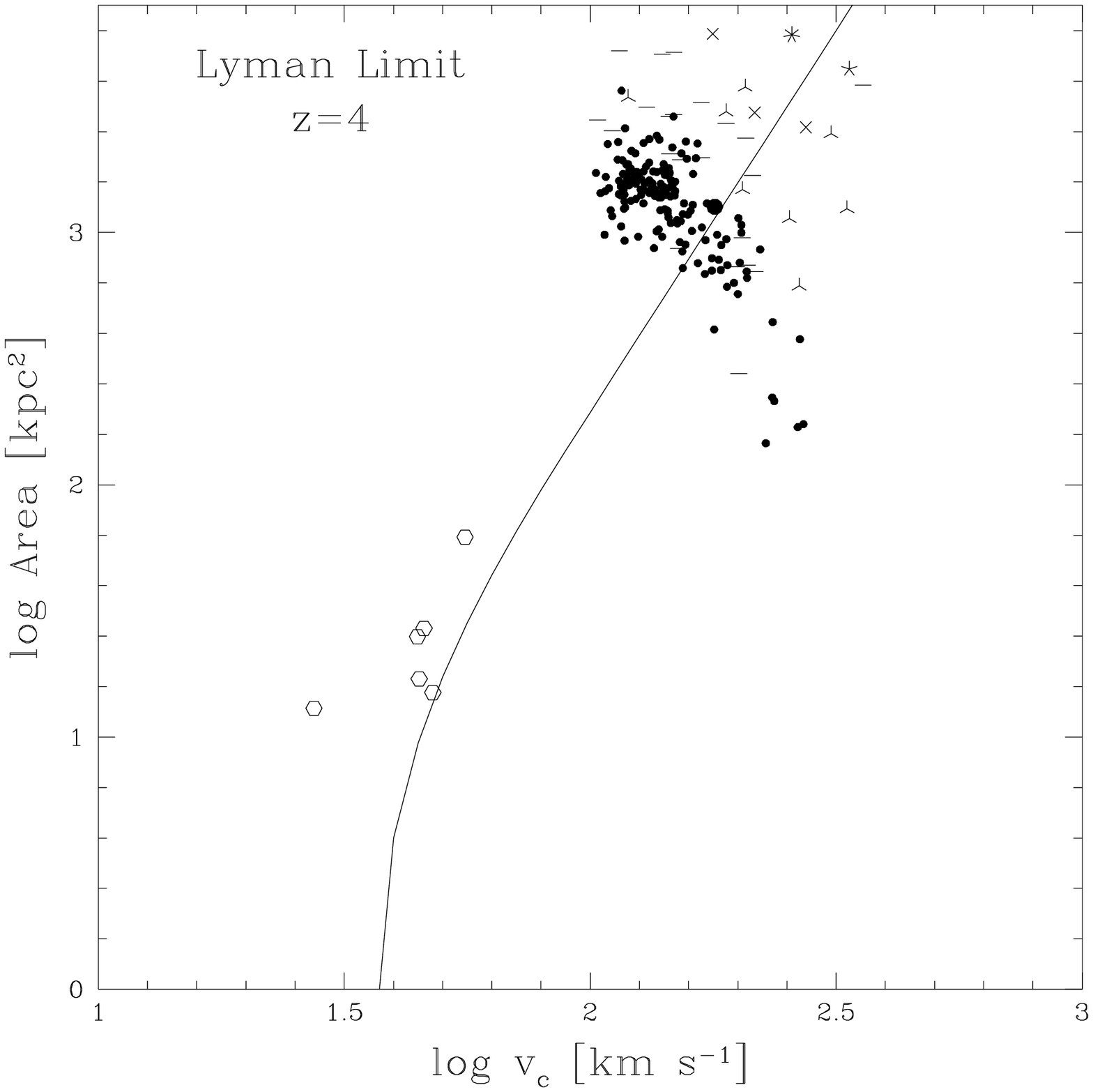}
\vglue-0.26in
\caption{Comoving absorbing area in kpc$^2$ vs. circular velocity
$v_c$ in \kmpersec\ for halos in the 22.22 Mpc simulation 
(skeletal points) and the 10 Mpc simulations (open circles).  
Left hand panels show the area for DLA absorption,
$\NHI \geq 10^{20.3}\;\cdunits$, and right hand panels
for Lyman limit absorption, $\NHI \geq 10^{17.2}\;\cdunits.$
The number of vertices in the skeletal points corresponds
to the number of gas concentrations in the halo.  The solid line shows the
fitted smooth relation of equation~(\ref{avc}), with
parameter values listed in Table 1.}
\label{figVAplot}
\end{figure}

Figure~\ref{figVAplot} shows the cross section for damped 
absorption (left hand panels) and
Lyman limit absorption (right hand panels) 
as a function of circular velocity for each of
our halos, at redshifts 2, 3, and 4.  
The open circles at low $v_c$ represent
halos from the 10 Mpc, high-resolution runs.  Other
points refer to the 22.22 Mpc simulation, and the number of vertices
in each symbol indicates the number of absorbers (i.e., distinct regions
of cold, collapsed gas) within each halo.  For these halos 
there are two competing effects that determine the trend between
absorption cross section and circular velocity.
Higher mass halos have deeper potential
wells, so concentrations of cold gas contract further, explaining the
downward trend in cross section with circular velocity exhibited by
points with a fixed number of vertices.  However,
more massive halos tend to harbor more than one
concentration of gas, increasing their absorption cross section.  
The overall trend in Figure 1 is that
halos of higher circular velocities on average have larger absorption
cross sections.  

The solid lines in Figure~\ref{figVAplot} show a smooth function
$\alpha_z(v_c)$ fitted to the relation between absorption area
and circular velocity.  We will need this function for our
Press-Schechter correction procedure below.  As a functional form
we adopt a linear relation between ${\rm log}\,\alpha$ and
${\rm log}\,v_c$ with a damping factor $1-\exp(-(v_c-35)/12)$,
which reflects the suppression of gas cooling in low $v_c$ halos.
We bin the data points in intervals of 0.15 in ${\rm log}\,v_c$,
compute the mean and standard deviation of ${\rm log}\,\alpha$
in each bin, and determine the parameters of the smooth
relation by $\chi^2$ minimization.  Fitting binned data rather
than individual halos gives more appropriate weight to the relatively
rare, high $v_c$ halos.  Table 1 lists the fitted values
of $A$ and $B$ for the functional relation
\be
{\rm log}\,\alpha = (A\,{\rm log}\,v_c + B)(1-\exp(-(v_c-35)/12)),
\label{avc}
\ee
with $\alpha$ in comoving kpc$^2$, $v_c$ in \kmpersec, and
base-10 logarithms.
We determine values separately for \dla\ and Lyman limit
absorption and for each redshift.  Figure~\ref{figVAplot}
shows that there is substantial scatter about this
mean relation, and our adopted functional form is rather arbitrary,
but we will see shortly that this characterization of the
$\alpha_z(v_c)$ relation suffices for our purposes.

\begin{table}
\begin{tabular}{lllll}
 \tableline\tableline
\multicolumn{1}{c}{$z$} & \multicolumn{1}{c}{$A_{\rm DLA}$} &
 \multicolumn{1}{c}{$B_{\rm DLA}$}& \multicolumn{1}{c}{$B_{\rm LL}$} &
 \multicolumn{1}{c}{$B_{\rm LL}$} \\ \tableline

2.0&   2.32&  -1.87  & 2.70  & -2.13 \\
3.0&   2.94&  -3.03  & 3.21  & -2.96 \\
4.0&   2.84&  -2.63  & 3.02  & -2.28 \\ \tableline\tableline

\end{tabular}
\caption{Fitted parameter values for $\alpha_z(v_c)$, with
the functional form in equation~(\ref{avc}).}
\label{tabalpha}
\end{table}

The observable quantity that we would like to test the CDM model
against is $n(z)$, the number of \dla\ or Lyman limit absorbers
per unit redshift interval along a random line of sight.
We can estimate this from the projected HI map of the 22.22 Mpc
simulation as in KWHM, by dividing the fractional area that has
projected column density above the \dla\ or Lyman limit threshold
by the depth of the box in redshift.  However, because the
simulation does not resolve gas cooling in halos with
$v_c \simlt 100$ \kmpersec, this procedure really yields
estimates of $n(z,100\;\vunits)$, where $n(z,v_c)$
denotes the number of absorbers per unit redshift produced
by halos with circular velocity greater than $v_c$.
Since halos with $35\;\vunits < v_c < 100\;\vunits$ can
harbor \dla\ and Lyman limit absorbers, 
$n(z,100\;\vunits)$ is only a lower limit to the observable 
quantity $n(z)$.

We have now assembled the tools to fix this problem, for the
Press-Schechter formula~(\ref{PSnumber}) tells us the number
density of halos as a function of circular velocity and the
relation $\alpha_z(v_c)$ tells us how much absorption these
halos produce.  Equation~(\ref{PSnumber}) is given in terms
of the mass $M$; since we define the halo mass within a sphere
of overdensity 177, the corresponding circular velocity is
\be
v_c = (GM/R_{177})^{1/2} = 
\left[GM^{2/3} \left({4\pi \over 3} 177 \rho_c\right)^{1/3}\right]^{1/2} =
117~ \left({M \over 10^{11} M_\odot}\right)^{1/3} 
\left({1+z \over 4}\right)^{1/2} \; \vunits.
\label{vcM}
\ee
Thus,
\be
	n(z,v_c)= {dr \over dz} \int_M^{\infty} N(M',z)
		\alpha_z(v_c) dM',
\label{nofzM}
\ee
where $N(M',z)$ is taken from equation~(\ref{PSnumber}), and 
equation~(\ref{vcM}) is used to convert between $v_c$ and $M$
as necessary.  Multiplying the comoving number density of halos by
the comoving absorption area yields a number of absorbers per
comoving distance, and multiplying by $dr/dz$, the derivative of
comoving distance with respect to redshift, converts to a number
per unit redshift.

\begin{figure}
\vglue-0.65in
\plottwo{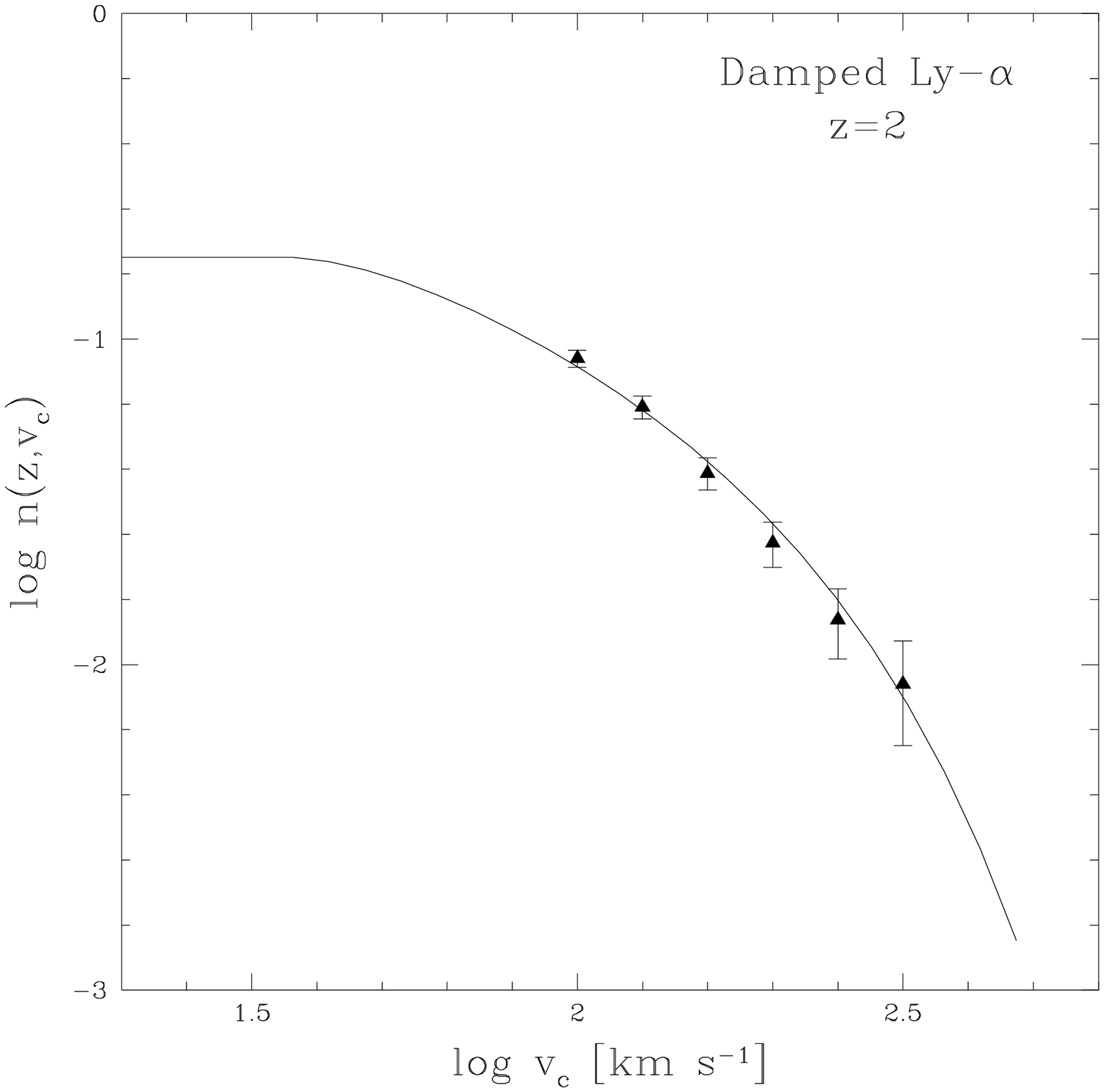}{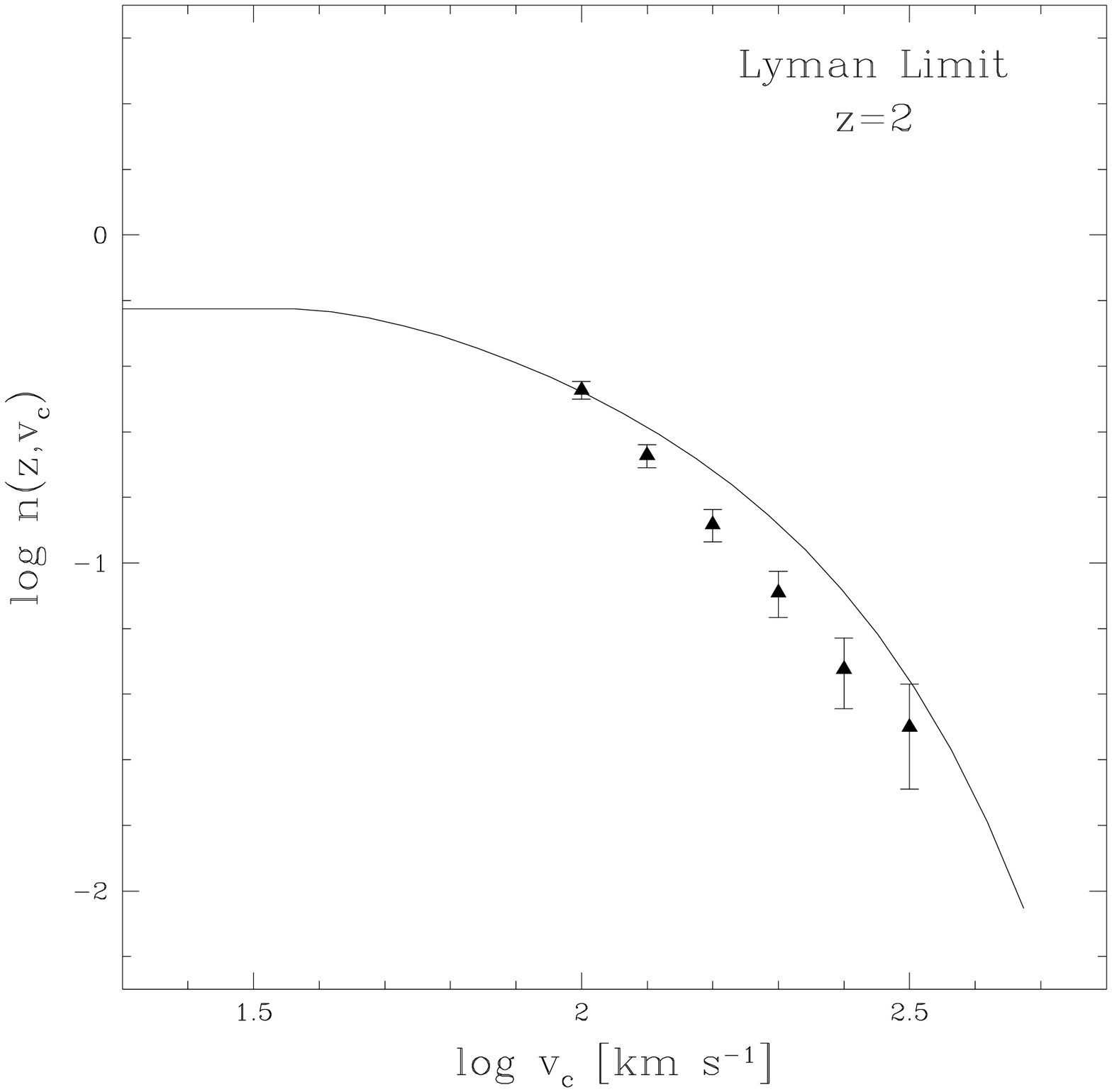} \\
\vglue-0.2in
\plottwo{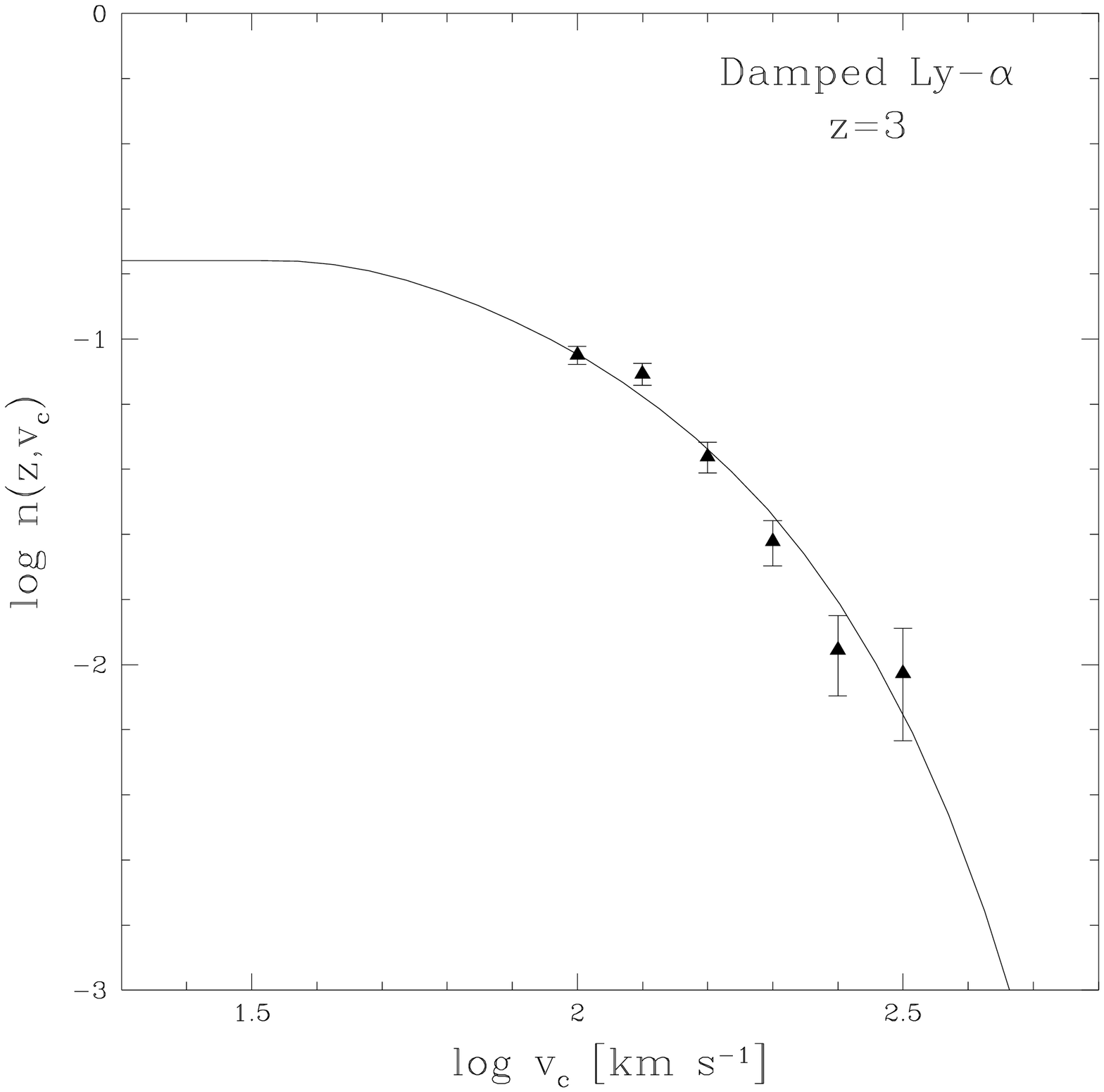}{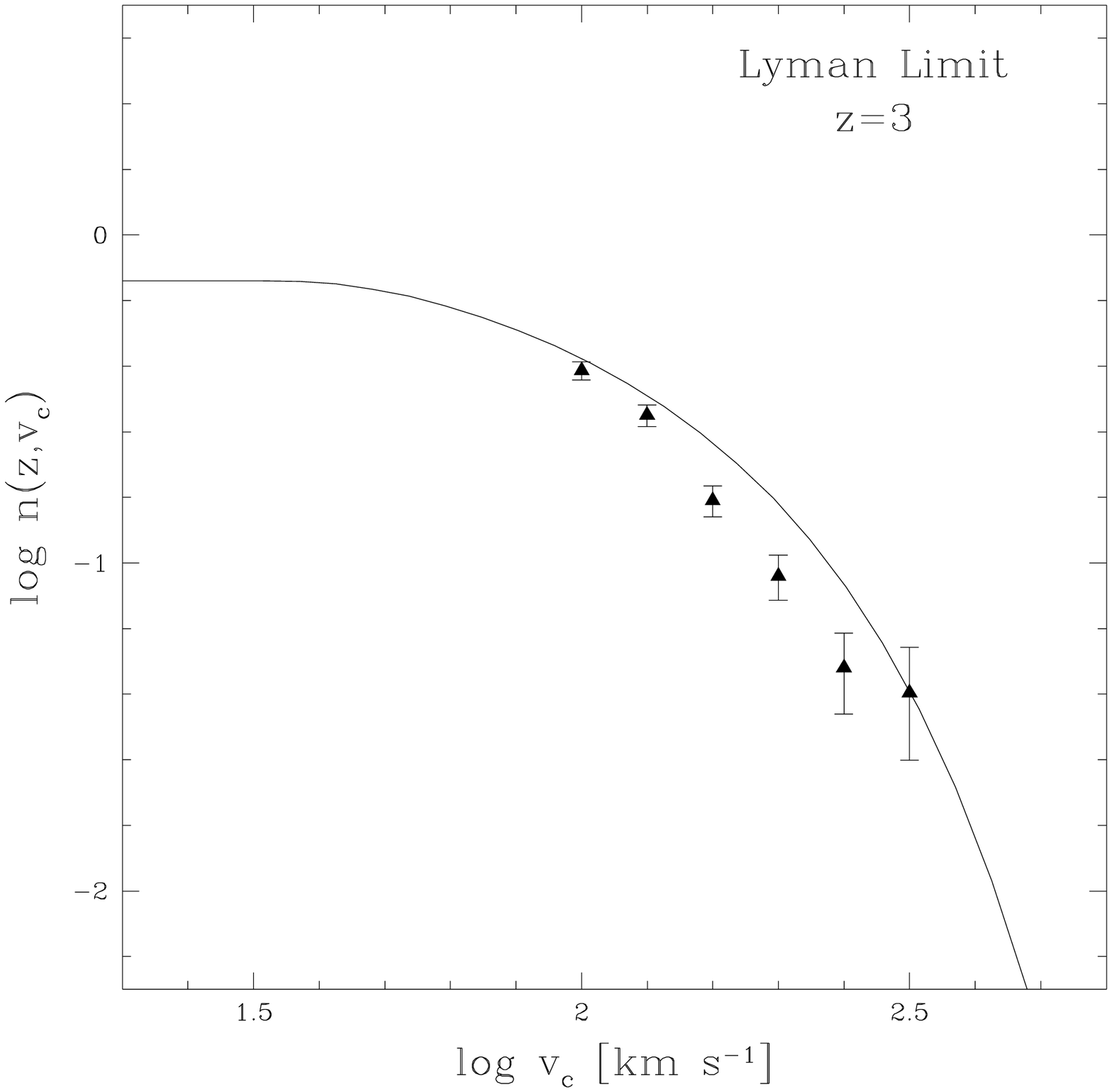} \\
\vglue-0.2in
\plottwo{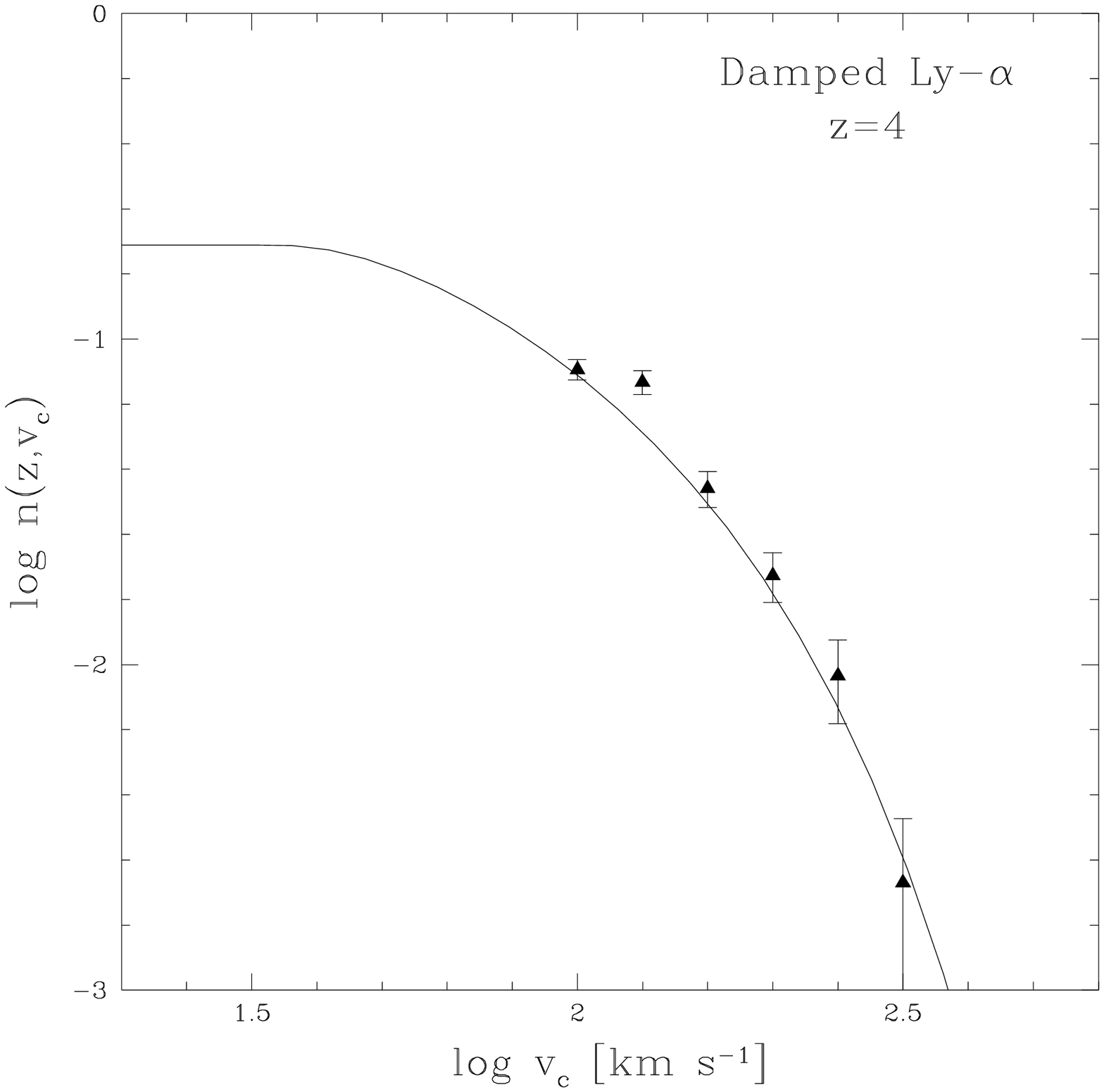}{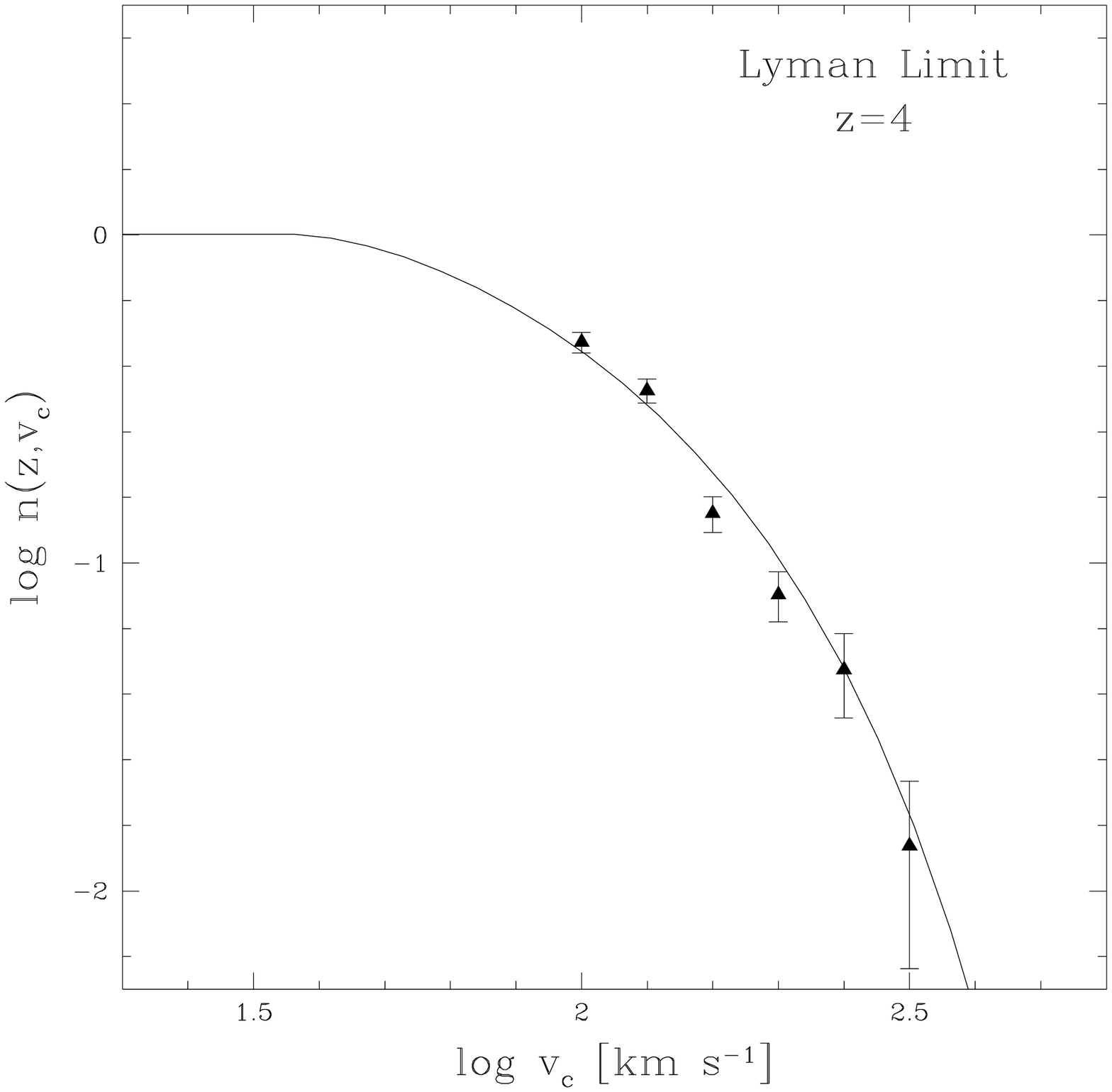}
\vglue-0.23in
\caption{Incidence of DLA (left) and Lyman limit (right)
absorption at $z=2,$ 3, and 4.  Curves show $n(z,v_c)$,
the number of absorbers per unit redshift arising in halos
with circular velocity greater than $v_c$, computed from
equation~(\ref{nofzM}).  The $y$-intercepts show the 
incidence of absorption produced by all halos.  Points with $N^{1/2}$ error
bars show numerical results from the 22.22 Mpc simulation.}
\label{figNZplot}
\end{figure}

Figure~\ref{figNZplot} shows $n(z,v_c)$ computed from 
equation~(\ref{nofzM}) using our fitted relations $\alpha_z(v_c)$.
Starting from high $v_c$, the abundance first rises steeply with
decreasing $v_c$ because of the increasing number of halos,
but it flattens at low $v_c$ because of the suppression of gas
cooling in small halos.  Points with error bars show $n(z,v_c)$
obtained directly from the halos in the 22.22 Mpc simulation.
The curves fit these points quite well --- they are, of course,
constructed to do so, but the agreement shows that our full
procedure, including the details of the Press-Schechter calibration
and fitting for $\alpha_z(v_c)$, is able to reproduce the original
numerical results in the regime where halos are resolved.
We can therefore be fairly confident in using this method to
extrapolate to $n(z,0) = n(z)$, the incidence of high column
density absorption produced by gas in all halos, thus
incorporating the new information provided by the high-resolution simulations.
These values of $n(z)$, the $y$-intercepts of the curves in
the panels of Figure~\ref{figNZplot}, are the principal numerical
results of this paper.  We will compare them to observations in
the next section.

Table 2 lists the values of $n(z)$ determined by this procedure
at $z=2$, 3, and 4.  It also lists the correction factors that
must be applied to the quantities $n(z,100\;\vunits)$ obtainable
by the KWHM procedure in order to get the total abundance
$n(z)=n(z,0)$.  In all cases, roughly half of the absorption 
occurs in halos with $v_c > 100\;\vunits$ and half in the
more common but smaller halos with lower circular velocities.

\section{Comparison to Observations}
\label{secResults}

\begin{table}
\begin{tabular}{lllcllclllcll}  \tableline\tableline
\multicolumn{6}{c}{Damped \lya} && \multicolumn{6}{c}{Lyman Limit}
\\ \cline{1-6} \cline{8-13}
\multicolumn{3}{c}{Calculated}&&\multicolumn{2}{c}{Observed}&&
\multicolumn{3}{c}{Calculated}&&\multicolumn{2}{c}{Observed}\\
z&\multicolumn{1}{c}{$n(z)$}&\multicolumn{1}{c}{$F_C$}&&
\multicolumn{1}{c}{$z$}&\multicolumn{1}{c}{$n(z)$} &&
z&\multicolumn{1}{c}{$n(z)$}&\multicolumn{1}{c}{$F_C$}&&
\multicolumn{1}{c}{$z$}&\multicolumn{1}{c}{$n(z)$}
\\ \cline{1-3}\cline{5-6}\cline{8-10}\cline{12-13}
2& 0.17857  &	2.05&&	$1.75\pm 0.25$&	$0.14\pm 0.073$ &&
2& 0.59586  &	1.74&&	$0.90\pm 0.5$&	$0.65\pm 0.25$ \\
3& 0.17411  &	1.91&&	$2.5\pm 0.5$&	$0.18\pm 0.039$ &&
3& 0.72439  &	1.81&&	$2.95\pm 0.6$&	$2.08\pm 0.35$ \\
 & & &&			$3.25\pm 0.25$&	$0.21\pm 0.10$ &&
 & & &&			&	 \\
4& 0.19422  &	2.54&&	$4.1\pm 0.6$&	$0.47\pm 0.17$ &&
4& 1.00660  &	2.31&&	$4.15\pm 0.6$&	$3.45\pm 0.95$ \\ \tableline\tableline
\end{tabular}
\caption{The incidence $n(z)$ of \dla\ and Lyman limit absorption for 
the $\Omega=1$ CDM model, computed by our calibrated
Press-Schechter procedure.  Observational values are taken from
Storrie-Lombardi \ea (1996) for \dla\ absorption and from 
Storrie-Lombardi \ea (1994) for Lyman limit absorption.
Also listed is $F_C$, the correction factor by which the KWHM results
for $n(z,100\;\vunits)$ must be multiplied to obtain the
absorption $n(z)$ produced by all halos.}
\label{tabResults}
\end{table}

\begin{figure}
\epsfxsize=6.5truein
\centerline{\epsfbox[18 144 590 718]{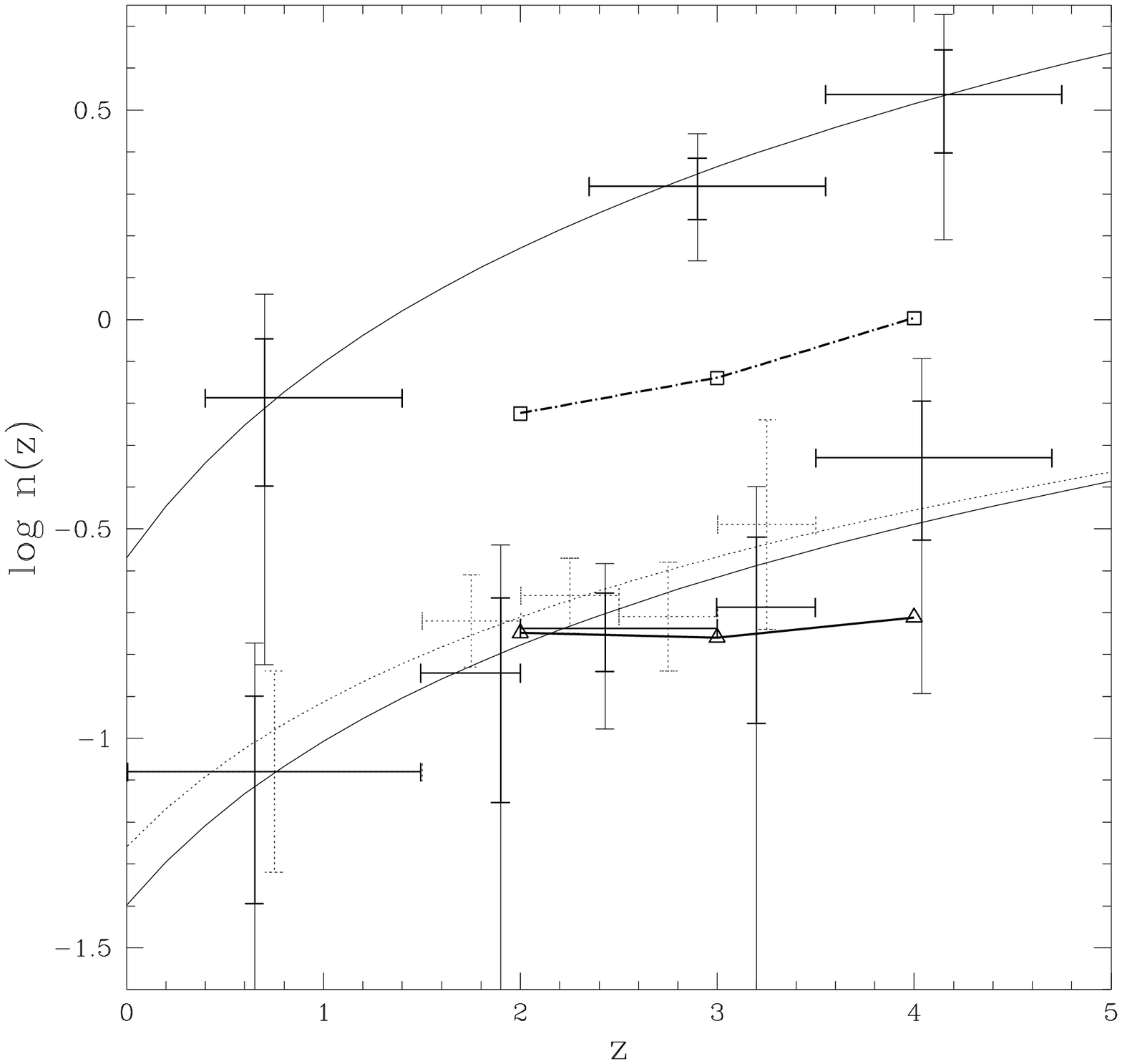}}
\caption{
\label{figObsComp}
Incidence of \dla\ and Lyman limit absorption as a function of
redshift.  Triangles and squares show the resolution-corrected
theoretical predictions for \dla\ and Lyman limit absorption,
respectively.  The upper error crosses represent the Lyman limit
data of Storrie-Lombardi \ea (1994), with $1\sigma$ and $2\sigma$
abundance errors shown.  The smooth curve shows their fitted power law.
The lower set of error crosses and solid curve represent the \dla\
data of Storrie-Lombardi \ea (1996), with $1\sigma$ and $2\sigma$
errors.  The dotted error crosses and curve show the data, $1\sigma$
errors, and fit from Wolfe \ea (1995).}
\end{figure}

Figure~\ref{figObsComp} compares our derived values of $n(z)$ to
observational estimates of the incidence of damped Ly$\alpha$
absorption, taken from Storrie-Lombardi \ea (1996) and
Wolfe \ea (1995), and Lyman limit absorption, taken from
Storrie-Lombardi \ea (1994).
The theoretical predictions and observed values are listed in Table 2.
The resolution correction increases the predicted $n(z)$ values
relative to those of KWHM by about a factor of two, leading to
quite good agreement with the observed abundance of \dla\ absorbers
at $z=2$ and 3.  At $z=4$ the predicted abundance is $1.6\sigma$
below the Storrie-Lombardi \ea (1996) data.  Since there are 
systematic as well as statistical uncertainties in this observational
estimate --- in particular, it includes candidate \dla\ systems
that have not yet been confirmed by Echelle spectroscopy ---
we regard this level of agreement as acceptable.

The situation for Lyman limit absorption is quite different.
Here the theoretical predictions fall systematically below the
observed abundances, by about a factor of three.
The correction for unresolved halos reduces the discrepancy found
by KWHM, but it does not remove it.
The deficit of Lyman limit systems could reflect a failing of
the CDM model considered here, or it could indicate the presence
in the real universe of an additional population of Lyman limit
absorbers that are not resolved by our simulations.
We discuss this issue further in \S~\ref{secSummary}

\section{Effects of Star Formation}
\label{secStars}

The simulations examined in the previous section do not allow
conversion of gas into stars, and one might worry that depletion
of the atomic gas supply by star formation would substantially
reduce the predicted abundance of \dla\ absorbers.
We investigate this issue by analyzing a simulation identical to the
KWHM run considered above except that it incorporates star formation.
The algorithm, a modified form of that introduced by Katz (1992),
is described in detail by KWH; we summarize it here.
A gas particle becomes ``eligible'' to form stars if (a)
the local hydrogen density exceeds 0.1 cm$^{-3}$ (similar to that of
neutral hydrogen clouds in the interstellar medium), (b)
the local overdensity exceeds the virial overdensity,
and (c) the particle resides in a converging flow that is Jeans-unstable.  
Star formation takes place
gradually, with a star formation rate that depends on an
assumed efficiency for conversion of gas into stars and on the local
collapse timescale (the maximum of the local dynamical timescale and
the local cooling timescale).  
We set the efficiency parameter defined by KWH to $c_*=0.1$,
though the tests in KWH show that results are insensitive to
an order-of-magnitude change in $c_*$.
Until the gas mass of such a particle
falls below 5\% of its original mass, it is categorized as a
``star-gas'' particle.  Thereafter, it is treated as a collisionless
star particle.  This gradual transition overcomes computational
difficulties associated with alternative implementations of
star formation, such as
the artificial reduction in resolution caused by rapid removal of
collisionless gas particles from converging flows, or the
spawning of a large number of extra particles that slow the
computations and consume memory.

When stars form, we add supernova feedback energy to the
surrounding gas in the form of heat, assuming that
each supernova yields $10^{51}$ ergs and that all stars greater than
$8M_\odot$ go supernova.
We add this energy gradually, with an exponential decay time of
2 $\times 10^7$ years, the approximate lifetime of an $8M_\odot$ star.
Thermal energy deposited in the dense, rapidly cooling gas is quickly radiated
away, so although feedback has some effect in our simulation, the
impact is usually not dramatic.

\begin{figure}
\epsfysize=5.0truein
\centerline{\epsfbox{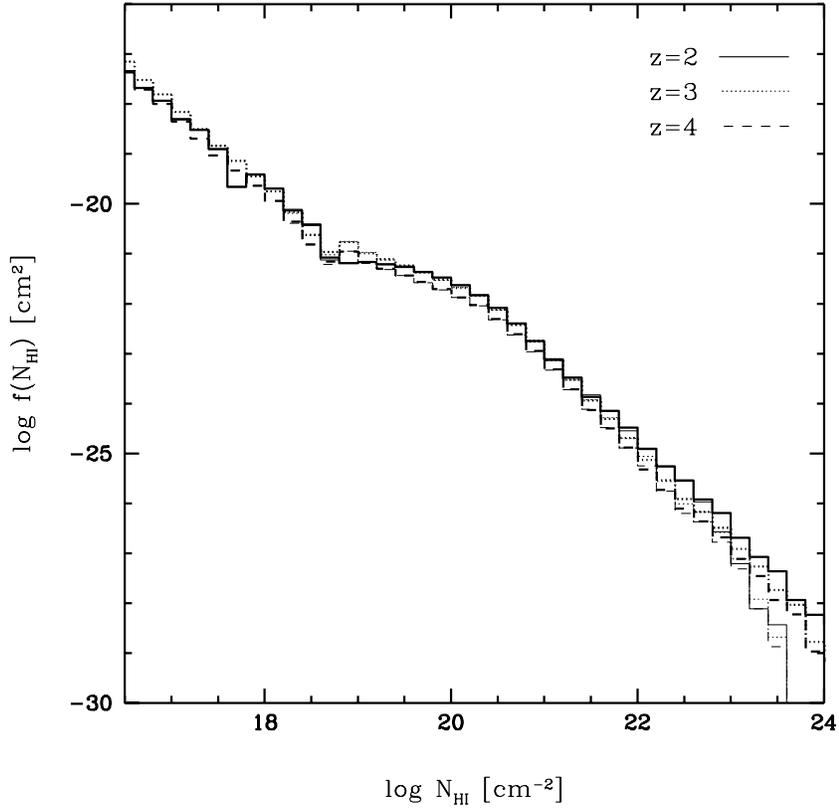}}
\caption{
\label{starfig}
The column density distribution $f(\NHI)$ --- the number of absorbers
per unit redshift per linear interval of $\NHI$ --- for simulations
with and without star formation.
Histograms show the simulation results at $z=2$ (solid), $z=3$ (dotted),
and $z=4$ (dashed).  Heavier lines represent the simulation
without star formation and lighter lines the simulation with
star formation.
}
\end{figure}

Figure~\ref{starfig} shows column density distributions
for the simulations with and without star formation at $z = 2$, 3, and 4;
$f(\NHI)$ is the number of absorbers per unit redshift per linear
interval of column density.  Star formation alters $f(\NHI)$ only at
column densities greater than $10^{22}$ cm$^{-2}$, higher than any observed
column density.
Star formation does affect the amount of cold, collapsed gas, however.  
The simulation without star formation
yields an $\Omega$ in cold, collapsed gas, i.e.\ gas with
$\rho/\bar\rho > 1000$ and $T<30,000$K, of (6.5, 3.6, 1.7)$\times 10^{-3}$
at $z = (2, 3, 4)$.  In the simulation with star formation,
the $\Omega$ in cold, collapsed gas is (3.4, 2.3, 1.2)$\times 10^{-3}$ 
at $z = (2, 3, 4)$, while the $\Omega$ in stars is 
(3.1, 1.2, 0.4)$\times 10^{-3}$, 
making a total $\Omega$ in collapsed galactic
baryons of (6.5, 3.5, 1.6)$\times 10^{-3}$, just slightly below the
simulation without star formation.  Hence, star formation simply
converts very high column density gas into stars while
affecting little else.
It does not significantly alter the predicted values of $n(z)$ given
previously because absorbers with $\NHI \geq 10^{22}\;\cdunits$
are a small fraction of all \dla\ absorbers.

All of the distributions in Figure~\ref{starfig} show a clear 
flattening in the column density range 
$10^{18.5}\;\cdunits \leq \NHI \leq 10^{20.5}\;\cdunits$.
This flattening reflects the onset of self-shielding.
A small range of total hydrogen column densities maps into a wider
range of neutral hydrogen column densities because the neutral
fraction rises rapidly with column density as self-shielding
becomes important.  While the optical depth to Lyman limit photons
is one at $\NHI = 10^{17.2}\;\cdunits$, self-shielding does not
become strong until significantly higher column densities because
higher frequency photons have a lower ionization cross section and
can still penetrate the cloud.

\section{Summary}
\label{secSummary}

The finite resolution of numerical simulations affects their predictions
for the abundance $n(z)$ of \dla\ and Lyman limit absorption systems.
It is not currently feasible to simulate a volume large enough to 
contain a representative population of high circular velocity halos
while maintaining enough resolution to accurately model the smallest
halos ($v_c \approx 35\;\vunits$) that can harbor such systems.
We have therefore devised a method that integrates results from high-
and low-resolution simulations to obtain accurate predictions for $n(z)$.
We use the simulations to determine the relation between absorption
cross section and halo circular velocity over the full range
of relevant circular velocities, then combine this relation with
the Press-Schechter formula for halo abundance --- itself calibrated
against the simulated halo population --- to compute $n(z)$ via
equation~(\ref{nofzM}).

As a method to correct for finite resolution, this technique should
be quite reliable, and it can be applied to other cosmological models
once the appropriate simulations are available for calibrating
$\alpha_z(v_c)$.  In the absence of these simulations, one can
make the plausible but uncertain assumption that the relation between
absorbing area and halo circular velocity is similar from one
model to another, then combine $\alpha_z(v_c)$ from this study
with the Press-Schechter halo abundance for other models to predict $n(z)$.
We apply this approach to a number of popular cosmological scenarios
in a separate paper (Gardner \ea 1996).
While it is less secure than the resolution-corrected numerical 
approach of this paper, it is an improvement over existing
semi-analytic calculations of \dla\ abundances
(e.g., Mo \& Miralda-Escud\'e 1994; Kauffmann \& Charlot 1994;
Ma \& Bertschinger 1994; Klypin \ea 1995),
which usually assume
that {\it all} gas within the halo virial radius cools and becomes neutral,
and which either assume a form and scale for the collapsed gas
distribution or compare to observations only through the
atomic gas density parameter $\Omega_g$, which is sensitive mainly
to the very highest column density systems.

Our resolution correction increases the incidence of
\dla\ and Lyman limit absorption in the CDM model by about a factor
of two, relative to the results of KWHM.  This increase brings the 
predicted abundance of \dla\ absorbers into quite good agreement
with observations at $z=2$ and 3, 
indicating that the high redshift galaxies that form in the CDM
model can account naturally for the observed damped \lya absorption.
At $z=4$ the predicted $n(z)$ is $1.6\sigma$ (a factor 2.4)
below a recent observational estimate.  However, many of
the systems that contribute to this data point have not yet been confirmed
by high-resolution spectroscopy, so the estimate may decrease with
future observations.

The underprediction of Lyman limit absorption in the simulations is
more dramatic, nearly a factor of three at $z=2$, 3, and 4.
This discrepancy could represent a
failure of the CDM model with our adopted parameters
($\Omega=1$, $h=0.5$, $\Omega_b=0.05$, $\sigma_8=0.7$),
though most of the popular alternatives to standard CDM have
less small scale power and therefore fare at least as badly in this regard.
An alternative possibility is that most Lyman limit
absorption occurs in structures far below the resolution scale of
even our high-resolution, individual object simulations.
For example, Mo \& Miralda-Escud\'e (1996) propose that most Lyman limit 
systems are low mass ($\sim 10^5 M_\odot$) clouds formed by thermal 
instabilities in galactic halo gas.
We could also be underestimating Lyman limit absorption if some of it
arises in partially collapsed structures --- sheets or filaments ---
that are not accounted for by the Press-Schechter halo formula.
While the KWHM simulation includes such structures, it may underestimate
their numbers in regions of low background density, where its spatial
resolution is degraded, and the QKE simulations select high density
regions from the outset.  High resolution
simulations focused on underdense regions could investigate this
possibility.  At lower redshifts Lyman limit absorption is
always associated with normal galaxies (Steidel \ea 1994; Lanzetta \ea 1996),
but this is not necessarily the case at high redshifts.

In addition to resolution-corrected estimates of $n(z)$, our
results provide some insights into the physical nature of \dla\ absorbers.
As shown in Figure~\ref{figNZplot}, roughly half of the absorbers
reside in halos with circular velocities greater than $100\;\vunits$
and half in halos with $35\;\vunits \leq v_c \leq 100\; \vunits$.
High resolution spectroscopy of metal-line absorption in damped
systems (e.g., Wolfe \ea 1994) may be able to test this prediction
over the next few years, and future simulations can provide predictions
for other cosmological models.  We find that halos with
$v_c \geq 150 \;\vunits$ frequently host more than one gas concentration
(Figure~\ref{figVAplot}), so imaging observations might often
reveal multiple objects close to the line of sight.

At $z\geq 2$, star formation and feedback --- at least as implemented
in our simulations --- have virtually no effect on the predicted
numbers of Lyman limit and \dla\ absorbers.  Roughly half of the
cold, collapsed gas is converted to stars by $z=2$, but this
affects the absorption statistics only at $\NHI \geq 10^{22} \;\cdunits$.
Depletion of the gas supply by star formation may account for the
absence of observed systems with column densities in this range,
though the number expected in existing surveys would be small in any case.
At lower redshifts, the effects of gas depletion may extend to lower 
column densities.  For $\Omega=1$ and $h=0.5$,
there are just over a billion years between $z=4$ and $z=2$,
but there are over two billion years between $z=2$ and $z=1$
and over eight billion years from $z=1$ to the present.
Assuming a roughly constant star formation rate in disk galaxies, 
most of the depletion of \dla\ gas would occur at low redshifts.

Ongoing searches for \dla\ absorbers are improving the observational
constraints on their abundance at high redshift, and
follow-up spectroscopic studies of their metal-line absorption
and imaging studies of associated \lya and continuum emission
are beginning to yield important insights into their physical
properties.  Multi-color searches for ``Lyman-break'' galaxies are
beginning to reveal the population of ``normal'' high redshift galaxies, 
which are the likely sources of most \dla\ absorption.
In the hierarchical clustering framework, the abundance,
properties, and clustering of these objects depend on the
amount of power in the primordial fluctuation spectrum on galactic
mass scales, which in turn depends on the nature of dark matter,
on the mechanism that produces the fluctuations, and on
cosmological parameters such as $\Omega$, $h$, and $\Omega_b$.
The initial fluctuations on galactic scales are difficult to 
constrain with local observations because much larger structures
(e.g., galaxy clusters) have since collapsed.  The comparison
between rapidly improving high redshift data and numerical
simulations like those used here opens a new window for testing
cosmological models, and we expect that it will take us much further
towards understanding the origin of quasar absorbers, high
redshift galaxies, and the galaxies that we observe today.

\acknowledgments

This work was supported in part by the San Diego, Pittsburgh, and Illinois
supercomputer centers, the Alfred P. Sloan Foundation, NASA Theory
Grants NAGW-2422, NAGW-2523, NAG5-2882, and NAG5-3111, NASA HPCC/ESS Grant 
NAG5-2213, NASA grant NAG5-1618, and the NSF under Grant ASC 93-18185
and the Presidential Faculty Fellows Program.

\vfill\eject
\end{document}